\documentclass[aps,prb,amsmath,amssymb,reprint]{revtex4-2}
\usepackage{graphicx}% Include figure files
\usepackage{dcolumn}% Align table columns on decimal point
\usepackage{bm}% bold math
\usepackage{subfigure}
\usepackage{graphicx}% Include figure files
\usepackage{dcolumn}% Align table columns on decimal point
\usepackage{bm}% bold math
\usepackage{braket}
\usepackage[utf8]{inputenc}
\usepackage[T1]{fontenc}
\usepackage{booktabs, array, mathptmx, float, tabularx, booktabs, lipsum, amsmath,multirow}
\usepackage{siunitx, xcolor}
\usepackage[version=4]{mhchem}

\begin{document}

\preprint{APS/123-QED}

\title{Geodynamics and artificial gravity\\ in space-time crystals under slow perturbation and deformation}
\date{\today}
%\thanks{}%
\author{Anzhuoer Li$^1$\protect\thanks{leean@}, Liang Dong$^2$\protect\thanks{liangdong@utexas}, and Qian Niu$^3$\protect\thanks{niuqian@ustc.edu.cn}}
\thanks{$^1$ leean@utexas.edu\\
$^2$ liangdong@utexas.edu
}
\affiliation{%
$^1$Department of Physics, University of Texas at Austin, Austin, TX 78712\\
$^2$University of Science and Technology of China, Hefei, Anhui, China\\
$^3$University of Washington, Seattle, USA
}%
\begin{abstract}
    We present a theory of geodynamics in a space-time crystal based on an event wave packet constructed from the Floquet-Bloch waves, which not only involve a scalar dispersion function but also a Berry curvature tensor in the phase space manifold of space-time and the reciprocal quasi energy-momentum.  In the presence of structural deformation, this theory is naturally extended into a covariant form with the introduction of a lattice connection constructed out of the gradients of the local lattice vectors. The geodesic equation for a particle not only involves the lattice connection but also higher-order corrections from space-time inhomogeneity of Berry curvatures and quasi energy-momentum dispersion gradients. These emergent connections and metric terms in the geodesic equations indicate the potential to experimentally realize artificial gravitational effects, thereby establishing a direct conceptual link between general relativity and quantum theory.
    \end{abstract}
\maketitle

 \section{Introduction} 
 Topological materials have attracted significant attention in recent years due to their rich and experimentally controllable phases. Among these, space-time crystals—materials exhibiting microscopic translational symmetry in both spatial and temporal dimensions \cite{rechtsman2013photonic,oka2019floquet}—are particularly intriguing due to their highly non-trivial topological characteristics \cite{zhoulongwen2025,shunyuyao2016,shunqingshen2024}. Considerable effort has been dedicated to the creation and investigation of such systems, motivated by their novel topological phenomena \cite{PhysRevB.106.224311,rudner2020band}. From a semi-classical viewpoint, these topological properties profoundly influence the motion of quasi-particles through geometric effects \cite{RevModPhys.82.1959}. Despite this, the dynamical behavior of quasi-particles in space-time crystals remains relatively unexplored. In our research, we focus specifically on the propagation of matter waves or excitations governed by linear partial differential equations with periodic dependencies in both position and time.

% A normal mode has the form of Floquet-Bloch wave, a plane wave with an amplitude varying periodically in space and time.   

In this paper, we seek a particle description by constructing a wave packet out of a narrow range of Floquet-Bloch waves of a single band, which is broad in real space compared with the space-time unit cell but still localized over macroscopic scales. This is the analog of a classical event point in the meta-space-time, and a set of geodynamic equations of motion for the corresponding worldline in phase space (space-time and the reciprocal quasi energy-momentum) can be obtained in the presence of weak external fields and various inhomogeneities. These equations are direct generalizations of equations of motion for electrons in three dimensional crystals. However, we treat space and time at equal footing and extend all the equations to phase space, and expect to construct artificial gravity through all emergent terms in geodesic lines.

We are also interested in the case of slowly varying deformation of the space-time crystal and wish to look into how such deformation affects particle dynamics. With the introduction of a lattice connection from the space-time lattice vectors and their reciprocals \cite{PhysRevResearch.2.013209}, a technique first introduced by two of us for the study of electron semiclassical dynamics in 3D crystals under deformation, we are able to obtain covariant geodynamics akin to Plebanski’s formulation of general relativity \cite{kleinert2005emerging}. 

The geodesic line is of the standard form involving the lattice connection, but both dispersion structure and Berry curvatures enter in higher order gradients indicating two ways of modification in effective space-time structure. There are also Lorentz force terms akin to gauge field effects and correcting the lattice connection.  

This paper is organized as follows: in section II, we present a general formalism to obtain Floquet-Bloch wave function in spacetime crystals. In section III, there are several direct and explicit examples presenting how the dispersions in different systems depend on the four-momentum. In section IV, an event wave packet is built with the Floquet-Bloch waves. The Lagrangian of the wave packet in phase space is presented, and the semi-classical equations of motion are derived. Section V tells the geodesic line equation based on the semi-classical equations. In section VI, we apply deformation in spacetime crystals and introduce lattice connections to indicate the inhomogeneity. Section VII and section VIII show how the equations of motion and geodesic equations are modified by deformation. We conclude in section VIII with a brief summary and outlook on our results.

\section{Geometro-wave equation and dispersion} 
The most general form of the linear equation governing quasi-particles evolution in spacetime crystals can be expressed using a 4-vector $x=(t,\vec{x})$,
\begin{equation}
    \hat{\mathcal{L}}(\frac{1}{i}\partial_x,x)\ket{\mathcal{\phi}}=0 \label{main 1}
,\end{equation}
where the $\ket{\mathcal{\phi}}$ is the wave function, and $\hat{\mathcal{L}}$ is the dispersion operator. Several examples of dispersion operators of varied quasi-particles are presented here: for Bloch electrons, the operator is $\hat{\mathcal{L}}=i\partial_t-\hat{H}(\frac{1}{i}\partial_{\vec{r}};\vec{x},t)$; for bosons like phonons in a crystal, the operator is $\hat{\mathcal{L}}=\partial^2_t-\partial_{\vec{r}}^2+m(\vec{x},t)$; and for a Dirac fermion, it is $\hat{\mathcal{L}}=iv_{F}^\mu \partial_\mu -m(\vec{x},t)$. 

Before we get into details of the dispersion operators, symmetry analysis offer important insights about the operators. All the operators in spacetime crystals obtain translational symmetry in space and time. The translational symmetry groups are space-time groups which are discrete \cite{PhysRevLett.120.096401}. Because of these symmetries, the operators have lattice structure , $ \hat{\mathcal{L}}(\vec{x},t)= \hat{\mathcal{L}}(\vec{x}+\vec{R}_\alpha,t+T_\alpha)$, from which we can define 4-dimensional lattice vectors as $c_\alpha=(T_\alpha,\vec{R}_\alpha)$.  According to the Floquet-Bloch theorem, when the wavefunctions follow the same periodicity in space and time with the corresponding operators, they can be written as product of a periodic part and a phase factor,
\begin{equation}
\ket{\mathcal{\phi}(k,x)}=e^{ik_\mu x^\mu}\ket{u(k,x)}  \label{main2}  
,\end{equation}
where $\ket{u(k,x)}$ are Floquet-Bloch functions, corresponding to the periodic part. The inner product of two Floquet-Bloch functions is integrated over a single lattice unit cell and a single period.
\begin{equation}
	\braket{u_1|u_2}=\frac{(2\pi)^4}{TV}\int_{V,T}d^4ru_1^* (x,k)u_2(x,k)
.\end{equation}

Equation \eqref{main 1} can be directly solved by substituting the equation \eqref{main 2},
\begin{equation}
	 \hat{\mathcal{L}}(\frac{1}{i}\partial_x+k,x)\ket{u(k,x)}=0
.\end{equation}
The results are $\omega=\omega(\vec{k})$, which are called dispersion relations. The dispersion relation is a hyper-surface in the phase space, but since we want to obtain wave packets with separation in the phase space, we may need to introduce a new method to break the constraint at first without loosing any necessary information.
 
 To obtain an independent 4-momentum, we introduce some new parameters to remove the dispersion constraint. We first interpret equation \eqref{main 1} as the zero "energy" solution of the following eigen-problem
\begin{equation}
      \hat{\mathcal{L}}(\frac{1}{i}\partial_x,x)\ket{\phi(k,x)}=\lambda \ket{\phi(k,x)} \label{maineq}
,\end{equation}
another parameter conjugated to $\lambda$, the "proper time", $\tau$ is also introduced. These two parameters could be viewed as energy and time generalized to four dimensions. By substituting a state varying with proper time $\tau$, $\ket{\Phi(k,x)}=e^{-i\lambda \tau}\ket{\mathcal{\phi}(k,x)}$, the steady equation of states is changed to proper time-dependent equation below, 
 \begin{equation}
      \hat{\mathcal{L}}(\frac{1}{i}\partial_x,x)\ket{\Phi(k,x)}=i\frac{d}{d\tau}\ket{\Phi(k,x)} \label{main 2}
,\end{equation}
and $\lambda$ is conserved to $\tau$. Although the real dispersion $\omega=\omega(\vec{k}) $ is dismissed and they are independent, the physical world lies in a shell $\lambda=0$. We need a four-dimensional spread momentum, so we have to go slightly off-shell with respect to the dispersion and allow components of the event wave packet to lie on a shell with small enough separation around $\lambda=0$, which causes no problem in the subsequent evolution of the wave packet because of the conservation law of $\lambda$.

The function $\lambda(k)$, known as the dispersion function or simply the dispersion, characterizes physical systems that typically reside near the shell defined by $\lambda(k) = 0$, which determines the allowed modes of propagation. To analyze the system near this shell, we expand $\lambda(k)$ in a Taylor series around a zero point $k_\mu = k_{0\mu}$: 
\begin{equation}
    \lambda(k) = m + \lambda^\mu (k_\mu - k_{0\mu}) + \frac{1}{2} \lambda^{\mu \nu} (k_\mu - k_{0\mu})(k_\nu - k_{0\nu}),
\end{equation}
where $m$ is an effective mass term, taken to be zero in this context but potentially dependent on space-time coordinates when deformations and perturbations are introduced. The coefficients $\lambda^\mu$ play a role analogous to 4-velocity, governing the linear behavior near the zero point, while the symmetric tensor $\lambda^{\mu \nu}$ captures second-order effects and, when non-degenerate, can be interpreted as a metric tensor that endows the dispersion surface with a geometric structure. By redefining the momentum as $q_\mu = k_\mu - \boldsymbol{A}_\mu$ to absorb the linear terms, the dispersion simplifies to a quadratic form, 
\begin{equation}
\lambda(q) = \tilde{m} + \frac{1}{2} \lambda^{\mu \nu} q_\mu q_\nu,
\end{equation}
which reveals the underlying geometric and dynamical properties of the system, particularly in contexts involving emergent space-time structure or quasi-relativistic behavior.

\section{dispersion examples}
The Floquet-Bloch function is periodic in space and time $u(k,x+c_\alpha)=u(k,x)$ because of Floquet-Bloch theorem. We can take the Fourier transformation of the Floquet-Bloch function, and use the exact diagonalization method to obtain the spectrum of dispersion. The Fourier transformation is,
\begin{equation}
    u(k,x)=\sum_n u_n \exp{( \mathbf{i} b^\alpha_{n\mu} x^\mu)},
\end{equation}
where $u_n$ are Fourier coefficients and $b^\alpha_n$ are the nth reciprocal lattice vectors. Substituting the Fourier series to the \eqref{main 2}, we can obtain the dispersion.

To indicate the dispersion spectrum, we select and analyze two representative examples. We focus our discussion on a single dispersion band for and present more replica bands of energy under on-shell constraint for simplicity. The first example is a one dimensional oblique Floquet crystal. In the Wannier basis, the operator is $\hat{\mathcal{L}}=i\partial_t-\hat{H}$ where the Hamiltonian is 
\begin{equation}
	\hat{H}=\sum_R [(V+A \cos ( \kappa R - \Omega t) ) \ket{R}\bra{R} + h(\ket{R+a} \bra{R}+ h.c )  ]
.\end{equation} 
Here, $R$ labels the lattice point with lattice constant $a$. The parameter $ \kappa$ and $  \Omega $ are the outer field's wave vector and frequency that changes the on-site energy $t$. $h$ is the hopping energy between two neighbor sites. The time translation symmetry is $H(t+2\pi/\Omega)=H(t)$. By Floquet-Bloch theorem \cite{PhysRevLett.128.186802} we can derive the spectrum of dispersion, which is the eigenvalue of the matrix below,
\begin{widetext}

\begin{align}
\left[
     \begin{matrix}
    \ddots\ & \cdots\ & \cdots\ & \cdots &\cdots \\
     \cdots  & \omega-\Omega-2h\cos((k-\kappa)\cdot a) & A/2 & 0 &\cdots\\
     \cdots & A/2 \ &\omega-2h\cos(k\cdot a) \ & A/2\ &\cdots \\
      \cdots & 0\ & A/2 \ &\omega+\Omega-2h\cos((k+\kappa)\cdot a) \ &\cdots \\
       \cdots\ & \cdots\ & \cdots\ & \cdots &\ddots.
    \end{matrix} \right]
\end{align}
\end{widetext}
 The figure of the dispersion is in FIG.1, and the energy bands $\omega=\epsilon(\vec{k})$ can be obtained by adding the constraint $\lambda=0$, which is also plotted in FIG.1.

Three points are highlighted in FIG.1, two are at the extrema (maximum and minimum) of a single dispersion relation band (point B and C) and one is in the middle (point A). At the three points in FIG. 1, we take the expansion mentioned before and find the spatial component $\lambda^{11}$ of effective metric $\lambda^{\mu \nu}$ is positive for point C and negative for point B. As for point A, since the leading order terms in both $k$ and $\omega$ are non-zero, linear terms are dominant. The non-zero expansion coefficients make the quasi-particle act like a massless Fermion. The different expansion parameters make quasi-particle centering in the region near A, B, and C transport in significantly different trajectories, as will be discussed in subsequent sections.

\begin{figure}[tbp] %% modification: fig 1 present index for \omega and k. different colors between zero values. oblique Brillouin zone.
	\centering
	\subfigure{
	%\begin{minipage}[c]{0.3\textwidth}
	
    \includegraphics[width=\linewidth]{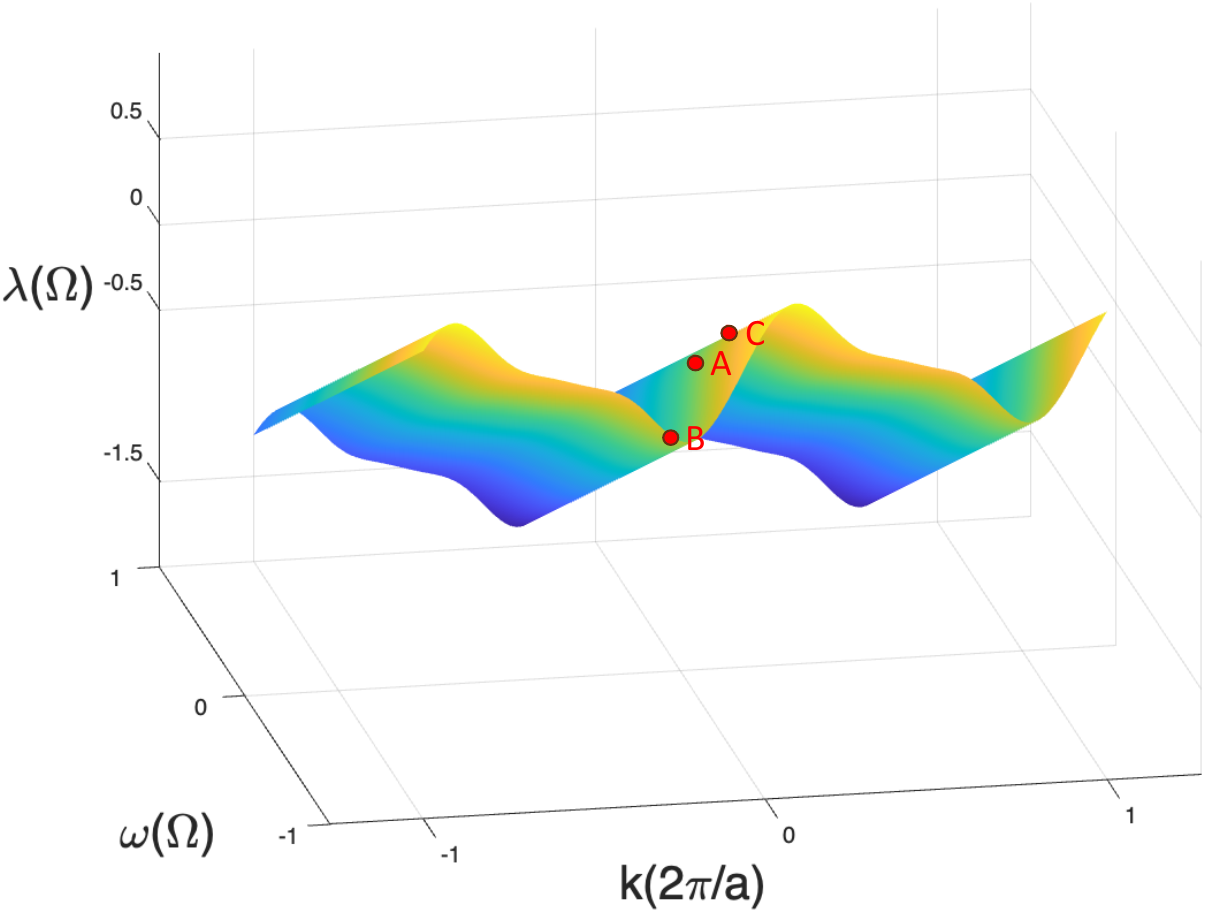}
    }
    \quad
    \subfigure{
    \includegraphics[width=\linewidth]{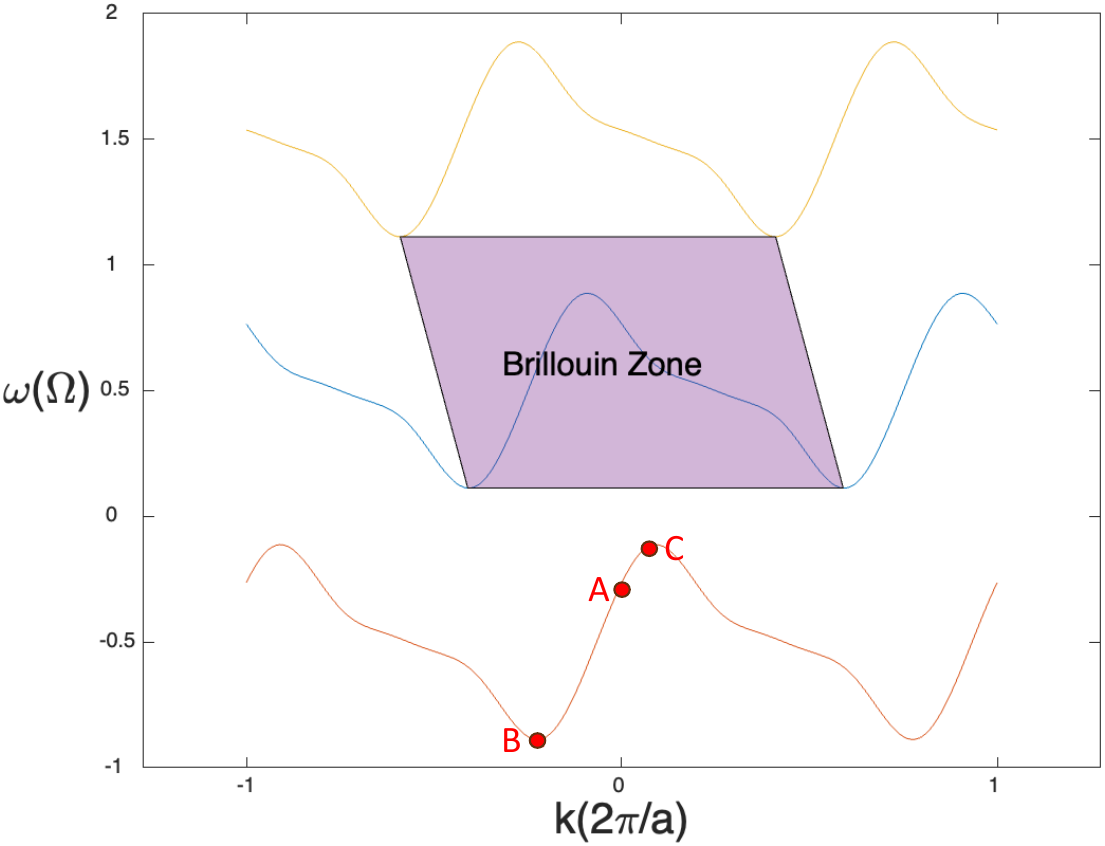}
    %\end{minipage}
    }
	\caption{The figure above is a single band dispersion in phase space and the figure below is dispersion relation $\omega=\omega(\vec{k})$. This single band spectrum of the oblique Floquet Bloch crystal is computed with parameter $h=0.4,t=0.5, A=1.5,\kappa=1.60$, here the parameters $h$, $t$, and $A$ are in units of $\Omega$ and $\kappa$ is in units of $2\pi/a$. $a$ is the lattice constant and $\Omega$ is the frequency of outer fields which are set to be 1. The dispersion relation $\omega=\omega(\vec{k})$ can be solved by taking $\lambda=0$. The oblique Brillouin zone is indicated by shaded region in the plot below, which is consistent with our previous analysis. }
\end{figure}

Another example is one-dimensional Floquet photonic crystal, which is common in Floquet system \cite{wang2021one}. The operator for 1-D photons with space-time translation symmetry is
\begin{equation}  %% equation of a gaped antiferromagnetic model. change to Dirac model.
	\hat{\mathcal{L}}=\partial_t^2-\partial_{r}^2 + A \cos (\Omega t - \kappa r)+M
.\end{equation}
The external field $ A \cos(\Omega t -  \kappa r)$ is determined by the wave vector $\kappa$ and frequency $\Omega$. There is an additional constraint for $M$ and $A$ is $M-A > 0$ to make the quasi-particles have positive mass. The dispersions are the eigenvalue of the matrix below, 

\begin{align}
\left[
     \begin{matrix}
    \ddots\ & \cdots\ & \cdots\ & \cdots &\cdots \\
     \cdots  \ & (\omega-\Omega)^2-(k-\kappa)^2 & A/2 & 0\ &\cdots\\
     \cdots & A/2 \ &\omega^2-k^2 \ & A/2\ &\cdots \\
      \cdots \ & 0\ & A/2 \ &(\omega+\Omega)^2-(k+\kappa)^2 \  &\cdots \\
       \cdots\ & \cdots\ & \cdots\ & \cdots &\ddots.
    \end{matrix} \right]
\end{align}

Two points are marked as D and E. These dispersion relations of these two points are hyperbolic curves in the $(\omega,k)$ plane, which are induced by the eigenvalues of the effective metric tensor $\lambda^{\mu \nu}$. The eigenvalues are one positive and one negative, and we can find such dispersions for particles moving on the space-like manifold. Like what we find in the first example, the parameters in expansion govern the motion of the quasi-particles.

%
%The spectrum for dispersion is below. For the 2 points in FIG. 2, the Fermi velocity and the metric term is 
%\begin{align}
%    \lambda^\mu_E=(-0.894,1.066) & \quad \quad \lambda^\mu_E=(0.029.625) \\
%    \lambda^{\mu\nu}_E= \left[
%    \begin{matrix}
%     & 15.78 \ &-4.54 \\

\begin{table}[htbp]
    \centering
    \begin{tabular}{|c|c|c|c|c|c|c|c|c|}
    \hline   
           & $k_1$ & $k_0$ & $\lambda^{1}$  & $\lambda^{0}$  & $\lambda^{11}$  & $\lambda^{10}$ & $\lambda^{01}$ & $\lambda^{00}$  \\
           \hline
        A & -0.03   & -0.47 & 7.47 &-1  &-8.68 & 0 &0  &0 \\
        B & -0.14 & -0.69  &0 & -1 & -152.39  & 0 &0 &0 \\
        C & 0.044  & -0.11 & 0 & -1 & 153.97  &0  &0 & 0\\
        D & 3.07  & -0.71 &0.62  &0.03  &0.20  &-6.88  &-6.88  &28.72 \\
        E & 1.04 & -1.21 &1.06  &-0.89  &1.64  &-9.10  &-9.10  &63.16 \\
        \hline
    \end{tabular}
    \caption{Expansion parameters of different points listed in the figure. 1. The first column is different points in figures, and the first row is different parameters. $k_{\mu}$ are zero points of $\lambda(k,\omega)$, $\lambda^{\mu}$ are linear expansion parameters, $\lambda^{\mu\nu}$ are quadratic parameters.}
    \label{tab:my_label}
\end{table}
%    &-4.54 \ &1.6375  
%    \end{matrix} \right] & \quad  \lambda^{\mu \nu}_D= \left[
 %   \begin{matrix}
 %    & 7.72 \ &-3.44 \\
 %   &-3.44 \ &0.20  
 %   \end{matrix} \right].
%\end{align}
%From the figure above, there are many Dirac cones (like points D and E) in the real energy bands, which indicates a method to engineer the band structure of Floquet crystal. There is an experiment showing that a spatially varying mass term may produce Dirac cones, which is consistent with our numerical results\cite{doi:10.1126/sciadv.abq4243}.

\begin{figure}[thbp]
	\centering
	\subfigure{
	%\begin{minipage}[htbp]{0.4\textwidth}
	\centering
	\includegraphics[width=\linewidth]{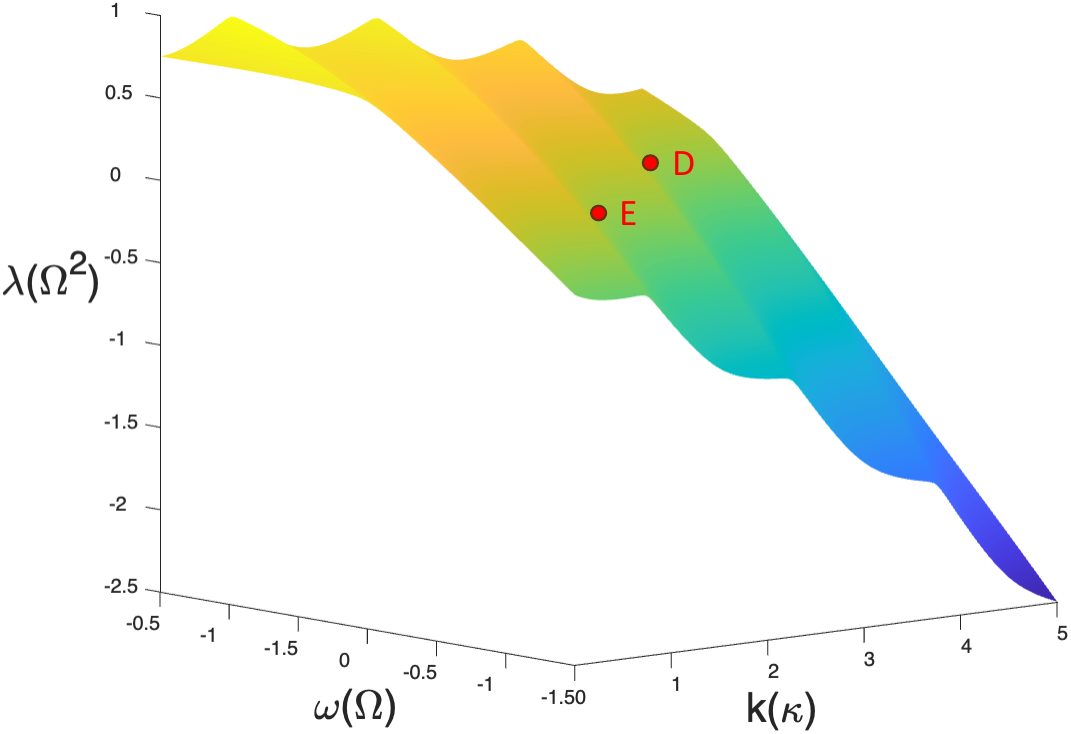}
	}
	\quad
	\subfigure{
	\includegraphics[width=\linewidth]{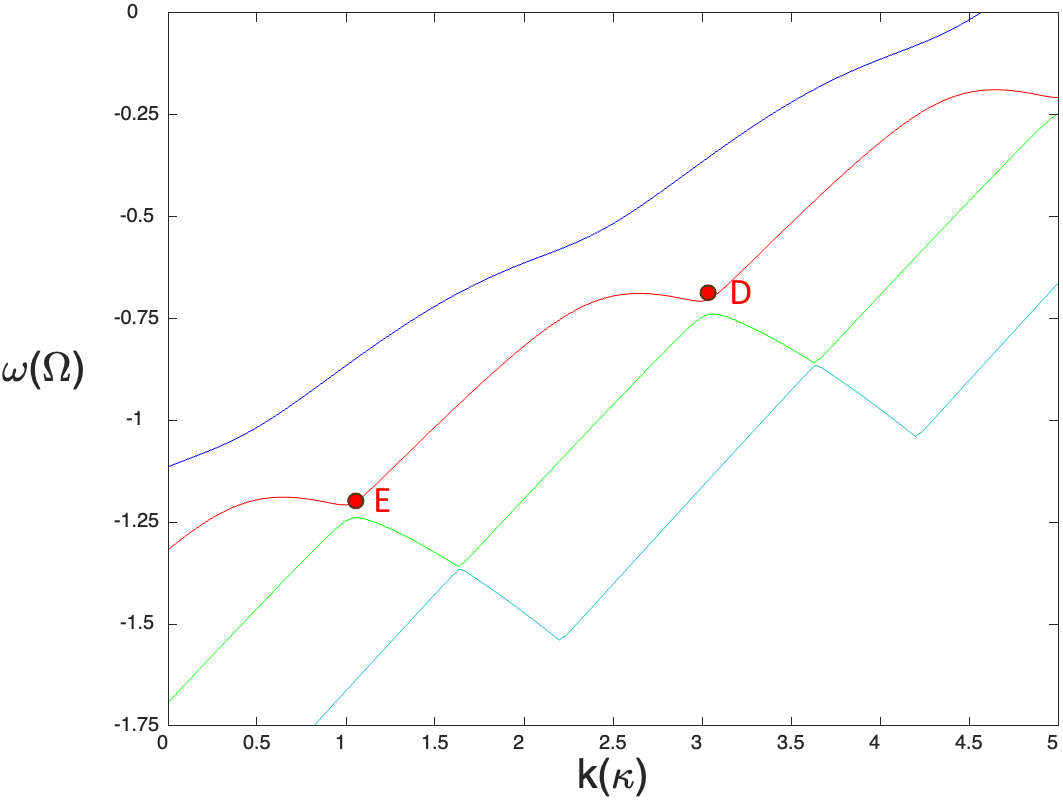}
    %\end{minipage}
	}
	\caption{Dispersion of boson with time translation symmetry. We select only one band to present. The parameters are $A=0.5,M=0.75$, where $k$ is in unit of $\kappa$ and $\Omega$, $A$, and $M$ are in units of $\Omega$. As for the $\lambda$ is in units of $\Omega^2$. The dispersion relation can be solved by taking $\lambda=0$. The figure below presents the dispersion of the band containing points D and E.}
\end{figure}

\section{event wave packet and geodynamics under slowly varying perturbation}
With the independent 4-momentum, the 4-dimensional wave packet can be constructed as
\begin{equation}
\textcolor{black}{\ket{W}=\int d^4 k\ a(k,\tau)e^{ik_\mu x^\mu} \ket{u(k,x)}}
,\end{equation}
where $\ket{W}$ is the wave packet state function and $a(k,\tau)$ is the distribution function localized in the center of the wave packet with normalization to make $\braket{W|W}=1$. The position of the center of the wave packet, where the wave packet is gathered in real space-time, can be calculated by the expectation value of the position operator $\hat{x}$ as $ \braket{W|\hat{x}|W}=x_c$. The momentum of the wave packet is also the expectation value from the function $a(k,\tau)$,
\begin{equation}
\textcolor{black}{k_c=\int d^4 k\ |{a(k,\tau)}|^2 k.}
\end{equation}

When dealing with a slowly varying potential whose characteristic length scale is significantly larger than the size of the wave packet, one can approximate the dispersion operator experienced by the wave packet at position $x_c$. Following the formulation in Ref.~\cite{sundaram1999wave}, this approximation involves adding a perturbation that explicitly depends on the wave packet center:
\begin{align}
\hat{\mathcal{L}} \approx \hat{\mathcal{L}}c + \sum_i (\hat{x}-x_c)^\mu \partial{x_c^\mu} \beta_i(x_c,t),
\end{align}
where the local dispersion operator $\hat{\mathcal{L}}_c = \hat{\mathcal{L}}(\hat{x}, \hat{p}; {\beta_i(x_c)})$ retains the space-time periodicity as if no perturbation were present. Consequently, equation \eqref{maineq} takes the modified form:
\begin{equation}
\hat{\mathcal{L}}\left(\frac{1}{i}\partial_x, x; x_c\right) \ket{\mathcal{\phi}(k,x; x_c)} = \lambda_c(k,x; x_c) \ket{\mathcal{\phi}(k,x; x_c)}.
\end{equation}
Here, the Floquet-Bloch functions and the corresponding dispersion explicitly depend on the space-time position $x_c$ of the wave packet. The adiabatic condition remains valid, given that these variations occur slowly compared to intrinsic timescales.

With the perturbation included, the dispersion relation \( \lambda = \bra{\mathcal{\phi}} \hat{\mathcal{L}} \ket{\mathcal{\phi}} \) can be evaluated up to first-order corrections using the following expansion:
\begin{equation}
\lambda \approx \bra{\phi} \hat{\mathcal{L}}_c \ket{\phi} + \bra{\phi} \Delta \hat{\mathcal{L}} \ket{\phi},
\end{equation}
where the gradient correction operator $\Delta \hat{\mathcal{L}}$ is given by
\begin{align}
\Delta \hat{\mathcal{L}} = \frac{1}{2}\left[(\hat{x}-x_c) \cdot \frac{\partial \hat{\mathcal{L}}}{\partial x} + \frac{\partial \hat{\mathcal{L}}}{\partial x} \cdot (\hat{x}-x_c)\right].
\end{align}
The expectation value of the local dispersion operator corresponds directly to the dispersion at the wave packet center:
\begin{equation}
\bra{\phi} \hat{\mathcal{L}}c \ket{\phi} = \lambda_c(x_c, k_c),
\end{equation}
and the gradient correction to the dispersion is expressed as
\begin{equation}
\Delta \lambda = -\text{Im}\left[\braket{\partial{x_c^\mu} u | \lambda_c - \hat{\mathcal{L}}c | \frac{\partial u}{\partial k\mu}}\right]\Big|_{k = k_c},
\end{equation}
which generalizes the gradient correction previously derived in Ref.~\cite{PhysRevB.59.14915} from three-dimensional to four-dimensional scenarios.

Due to the presence of space-time inhomogeneity, Berry connections associated with space-time coordinates do not vanish. They are defined as follows:
\begin{align}
\begin{split}
A_{x^\mu}=i\braket{u(k,x)|\partial_{x^\mu}|u(k,x)}, \\
\textcolor{black}{A_{k_\mu}=i\braket{u(k,x)|\partial_{k_\mu}|u(k,x)}.}
\end{split}
\end{align}
The total Berry curvatures are introduced in the eight-dimensional phase space by differentiating these Berry connections with respect to the four-momentum and four-coordinates:
\begin{gather}
\begin{split}
\Omega_{k_\mu k_\nu}= i(\braket{\partial_{k_\mu} u| \partial_{k_\nu} u}-\braket{\partial_{k_\nu} u | \partial_{k_\mu} u} ),\\
\Omega_{k_\mu x^\nu}= i(\braket{\partial_{k_\mu} u | \partial_{x^\nu} u}-\braket{\partial_{x^\mu} u | \partial_{k_\nu} u}),\\
\Omega_{x^\mu x^\nu}=i(\braket{\partial_{x^\mu} u |\partial_{x^\nu} u } -\braket{\partial_{x^\nu} u|\partial_{x^\mu} u}).
\end{split}
\end{gather}

The semi-classical action for the particle-like wave packet is given by $S = \int L \, d\tau$, where the Lagrangian $L$ is defined as $L = \braket{W|i\frac{d}{d\tau}-\hat{\mathcal{L}}|W}$. Explicitly, the Lagrangian can be expressed as \cite{sundaram1999wave}
\begin{equation}
     L=-\lambda + k_{c\mu} \dot{x}_{c}^\mu + \dot{k}_{c\mu} A_{k_{c\mu}} +\dot{x}_{c}^\mu A_{{x_c}^\mu}. \label{lag}
\end{equation}
This formulation generalizes the three-dimensional lattice scenario, incorporating the dependence on wave packet center positions, to an extended description applicable to a phase space with variational corrections in the dispersion operator.

Using the variation principle, the corresponding equations of motion can be derived:
\begin{gather}
        \dot{ x}^\mu = \frac{\partial \lambda}{ \partial k_{ \mu} }|_{\lambda=0} - \Omega_{k_{\mu } x^\nu} \dot{x^\nu} - \Omega_{k_{\mu } k_{\nu }} \dot{k}_{\nu}, \\
\dot{k}_{\mu} = \frac{\partial\lambda}{\partial x^\mu}|_{\lambda=0} + \Omega_{x^\mu x^\nu }\dot{x^\nu}+\Omega_{x^\mu k_{\nu}} \dot{k}_{\nu}, \label{main 8}
    \end{gather}
where $\dot{x}^\mu = \frac{dx^\mu}{d\tau}$ and $\dot{k}_{\mu} = \frac{dk_{\mu}}{d\tau}$. In subsequent discussions, since the physical results assume the condition $\lambda = 0$, this constraint will not be explicitly stated. An alternative method of presenting this constraint using an auxiliary field is explored in the Appendix.

\section{ Geodesic equations and artificial gravity}
The phase‐space equation of motion \eqref{main 4} can be viewed as describing particles propagating on a four‐dimensional manifold, with extra “force” terms arising from the underlying dispersion band structure and geometry.  A key question is whether these additional couplings genuinely modify the effective space-time manifold through which the particles move.  In what follows, we will demonstrate that the dispersion derivative terms and Berry‐curvature contributions are effectively in the form of space-time connections and gauge fields.

To illustrate the dynamics clearly, we begin by neglecting all Berry curvatures, focusing solely on the pure dynamical effect of wave packet motion. Under this simplification, the equations of motion reduce to:
\begin{gather}
\begin{split}
   \dot{x}^\mu &= \frac{\partial \lambda}{ \partial k_{ \mu} } \label{main_flat 1} \\ 
          \dot{ k}_{\mu} =& -\partial_{x^\mu}\lambda.
\end{split}
\end{gather}

To derive the geodesic equation, which reveals how the underlying space-time manifold is curved, we need an equation dependent solely on coordinates and their derivatives. By differentiating the coordinate component once more and substituting the momentum expression, we obtain the second-order equation:
\begin{equation}
    \ddot{x}^\mu =\dot{ k}_{\nu} \partial_{k_{\mu}} \dot{x}^\nu -\dot{x}^\nu \partial_{k_{\mu}} \dot{k}_{\nu} \label{eqflat 1}
.\end{equation}
The left-hand side represents the acceleration, while the right-hand side can be interpreted as a rotation-like term in phase space.

For the sake of later discussion, we introduce the concept of a free particle here. For a free particle, no external force arises from spatial or temporal variations in dispersion, implying:
\begin{equation}
\partial_{x^\mu}\lambda(k,x)=0.
\end{equation}
Moreover, this condition remains valid around the momentum center of the wave packet. As a free particle propagates, it experiences no force induced by derivatives of dispersion in momentum space, leading to:
\begin{equation}
\partial_{k_\nu}\partial_{x^\mu}\lambda(k,x)=0.
\end{equation}
Under these conditions, the equations of motion simplify significantly:
\begin{equation}
\ddot{x}^\mu=0. \label{flat_geo_1}
\end{equation}
This result confirms that the acceleration of a free particle must vanish, aligning with a generalized form of Newton's second law.

For a more general discussion, we may expand the dispersion relation around a zero point to various orders. Focusing specifically on neighborhoods of zero points with nonzero first-order derivatives, such as point A in Fig. 1, the dispersion can be expressed in a linear form as $\lambda = \lambda^\mu k_\mu + \lambda_0$. Here, the Fermi velocity $\lambda^\mu$ and the deformation potential $\lambda_0$ depend solely on the spatial coordinates $x$. A similar dispersion relationship is found near Dirac points in Floquet graphene \cite{PhysRevB.90.115423, PhysRevLett.107.216601}. Under these conditions, the geodesic equation simplifies to
\begin{equation}
\ddot{x}^\mu = -\dot{x}^\nu \partial_{x^\nu}\lambda^\mu,
\end{equation}
illustrating explicitly how the acceleration is influenced by both velocity and spatial variations.

When quadratic terms are taken into account, the metric terms $\lambda^{\mu \nu}$ lead to modifications in the effective space-time structure. To simplify the quadratic expansion, we introduce a shifted momentum $q_\mu = k_\mu - \boldsymbol{A}_\mu$. Substituting this shift, the dispersion relation becomes:
\begin{equation}
\begin{aligned}
\ddot{x}^\mu+\tilde {\Gamma}^\mu_{\rho\sigma}  \dot{x}^\rho \dot{x}^\sigma= \lambda^{\mu \nu} \partial_\nu \tilde{m}-\lambda^{\mu\nu} \boldsymbol{F}_{\nu \rho} \dot{x}^\rho, \label{flat_geo 2} \\
\tilde{ \Gamma}^\mu_{\rho\sigma}= \frac{1}{2} \lambda^{\mu \beta}(\partial_\beta \lambda_{\rho \sigma} -\partial_{\rho} \lambda_{\beta \sigma}- \partial_{\sigma} \lambda_{\beta \rho}).
\end{aligned}
\end{equation}
Here, the new connections $\tilde{\Gamma}^\mu_{\rho\sigma}$, which serve as Levi-Civita symbols induced  from the effective metric $\tilde{g}_{\mu\nu} = \lambda_{\mu\nu}$, explicitly link the space-time geometry to the dispersion relation. Additionally, the emergent Abelian gauge fields $\boldsymbol{F}_{\nu\rho} = \partial_\nu\boldsymbol{A}_\rho - \partial_\rho\boldsymbol{A}_\nu$ produce a term analogous to the Lorentz force.

Now we add berry curvatures back and keep the free particle constraint. The equations of motion \eqref{main 8} may be simplified to the form below
\begin{equation}
    (\delta_\mu^\nu -\Omega_{x^\mu k_{\nu}}) \dot{k}_{\nu} = \Omega_{x^\mu x^\nu }\dot{x^\nu}. \label{main 9}
\end{equation}
The equations above show that the particle is moving under the artificial gauge field, which could induce topological Hall effect \cite{bruno2004topological}. As for the spatial part,  they are 
\begin{equation}
    \dot{ x}^\mu = \frac{\partial \lambda}{ \partial k_{ \mu} } - \Omega_{k_{\mu } x^\nu} \dot{x^\nu} - \Omega_{k_{\mu } k_{\nu }} \dot{k}_{\nu} 
.\end{equation}

Taking an additional derivative of the coordinate equation, the geodesic equation can be further refined to:
\begin{equation}
\begin{aligned}
\ddot{x}^\rho +& \partial_{x^\sigma}\Omega_{k_\rho x^\nu}\dot{x}^\sigma\dot{x}^\nu =\\& \left(\Omega_{x^l x^\alpha}\partial_{k_\alpha}\partial_{k_\rho}\lambda + \Omega_{k_\rho k_\nu}\partial_{x^l}\partial_{x^\nu}\lambda\right)\dot{x}^l. \label{flat_geo_3}
\end{aligned}
\end{equation}
Equation \eqref{flat_geo_3} is a second-order correction of Eq. \eqref{flat_geo_1}, incorporating the effects induced by gauge fields. The terms $\partial_{k_\alpha}\partial_{k_\rho}\lambda$ and $\partial_{x^l}\partial_{x^\nu}\lambda$ act as effective metrics in phase space that couple to the dual Berry curvatures. Berry curvatures with indices in real space ($\Omega_{x^l x^\alpha}$) or momentum space ($\Omega_{k_\rho k_\nu}$) are similar to electromagnetic fields in general relativity \cite{carroll2019spacetime}. Additionally, the Berry curvature $\Omega_{k_\rho x^\nu}$ modifies the density of states in phase space in the absence of external fields \cite{PhysRevLett.95.137204}. Its spatial covariant derivative contributes directly to alterations in wave packet trajectories, demonstrating the impact of phase-space dynamics on real-space motion.

\section{geometry of deformation in space-time crystal}
In addition to perturbations induced by the wave packet center, deformations of the crystal structure also play a crucial role. Spacetime deformations, easily controllable in certain cold-atom systems, significantly affect the motion of quasi-particles. Extending our description from a flat spacetime crystal to a deformed spacetime crystal necessitates specifying local crystal variations explicitly. Consequently, introducing a lattice connection becomes essential to characterize these local structural changes precisely.

Space-time lattice connection demonstrates the inhomogeneity of spacetime crystals. When there is deformation in space-time crystal, the lattice vectors become vector fields varying with the position and time, which means  $c_\alpha=c_\alpha(x)$. The form is a direct generalization of local lattice connection in three-dimensional deformed lattices \cite{PhysRevB.98.115162,liangdongthesis}. Because of the orthogonal condition between lattice vectors and reciprocal lattice vectors, the reciprocal lattice vectors are also vector fields $b^\alpha=b^\alpha(x)$. To derive the form of lattice connection, consider the lattice vectors change
$d c_\alpha^\mu=\partial_\nu  c_\alpha^\mu dx^\nu$, we define the lattice connection to make the change of lattice vector fields proportional to the product of lattice connection and lattice vector fields, 
\begin{equation}
    dc_\alpha^\mu=\Gamma^{\mu}_{\sigma m} c_{\alpha}^\sigma dx^m
,\end{equation}
where the $\Gamma^{\mu}_{\sigma m}$ is the connection. By substituting reciprocal lattice vectors,
\begin{equation}
    \Gamma^{\mu}_{\sigma m}=b_\sigma^\alpha \partial_m c_\alpha^\mu
.\end{equation}
The adiabatic condition of deformation requires a relative small deformation, so all the derivatives of lattice vector fields are small terms $\left|\frac{\partial_\mu c_{\alpha}(x)}{c_{\alpha(x)}}\right| \textless \textless1$. In momentum space, all the reciprocal lattice vectors are also slowly deformed and satisfy $\left|\frac{\partial_\mu b^{\alpha}(x)}{b^{\alpha}(x)}\right| \textless \textless1$ .

Under deformation that is conspicuous only in large scale space time length, the lattice potential no longer maintains strict periodicity. However, it retains local periodicity within small space-time regions. Consider the one-dimensional boson model as an example, where the lattice potential is expressed as $A \cos(\Omega t - \kappa r)$. Under deformation, we set $\Omega(x,t) = 5 + 0.1 \cos(x)$, $\kappa(x,t) = 5 + 0.1 \sin(x)$, and $a(x,t) = 0.01 \sin(x)$. All parameters here are dimensionless. Below, we illustrate the potential over an extended range to highlight its deformation clearly, and we select a smaller region to demonstrate that the potential experienced by a wave packet locally still obtains translational symmetry approximately.

 % needs additional modification
\begin{figure}[tbp]
	\centering
    \subfigure{
    \centering
	\includegraphics[scale=0.2]{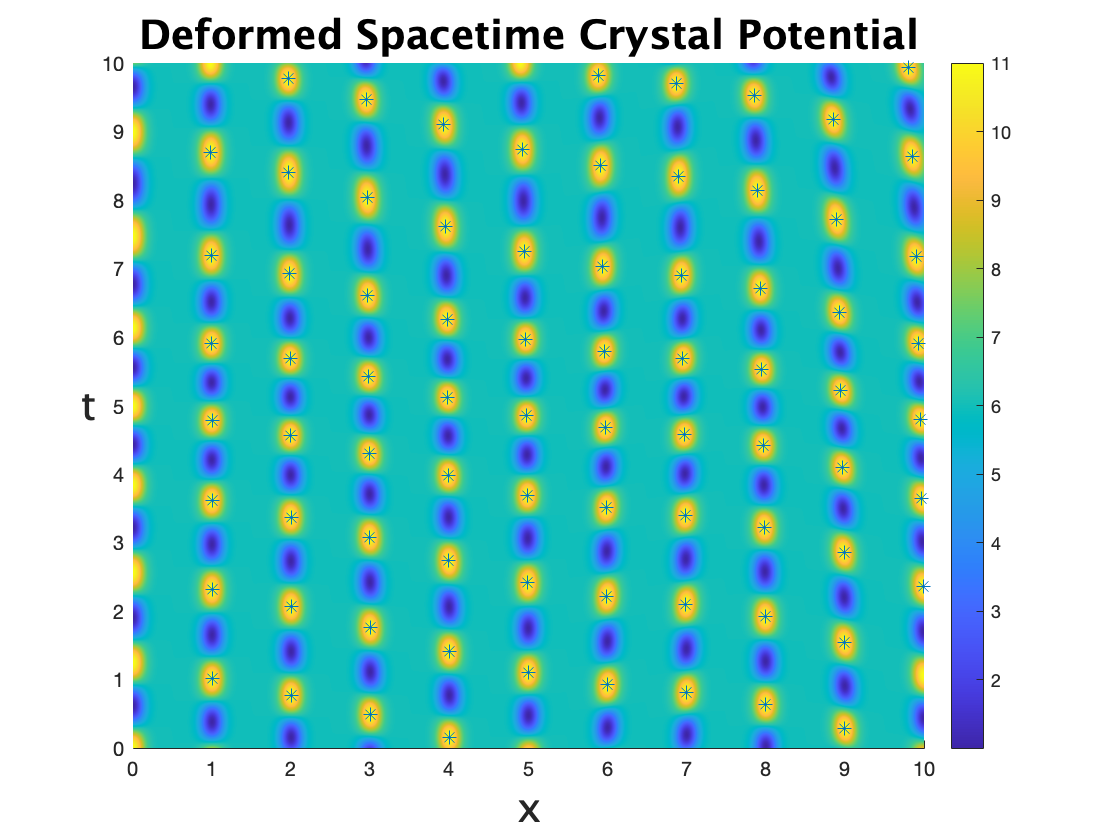}
    }
    \quad
    \subfigure{
    \centering
    \includegraphics[scale=0.2]{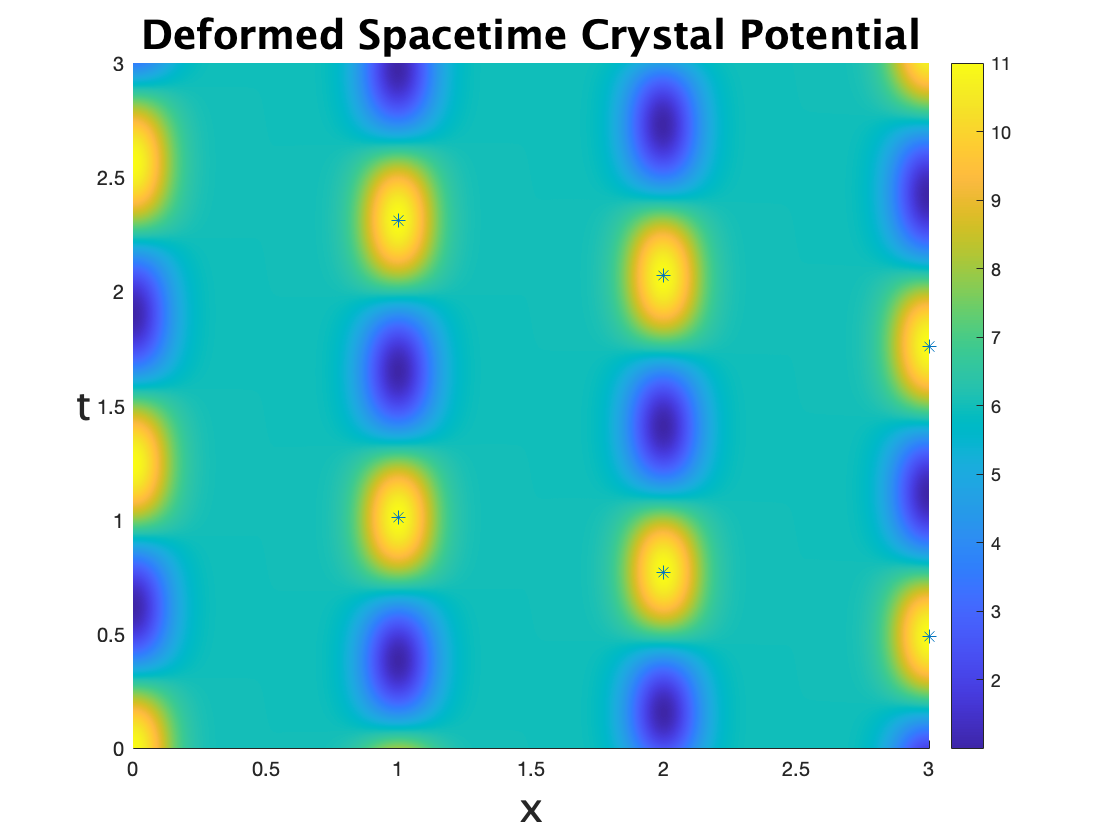}
    }
	\caption{\textcolor{black}{Graphs of the lattice potential of 1-D boson space-time crystal. The parameter is $\Omega=5+0.1\cos(x), \kappa=5+0.1\sin(x), a=0.01\sin(x)$. The maximum point of lattice potential is marked by a star label. The figure above is the potential over a large space-time zone, where significant variation of periodicity can be found. The figure below is the potential in a small regime that potential form almost a constant lattice structure}}
\end{figure}

The lattice connection defined earlier may exhibit an antisymmetric component due to the discrete nature of the space-time lattice points. Nonetheless, wave packet dynamics remain valid for describing localized space-time wave packets provided there are no topological defects in the space-time crystal. The presence of topological defects introduces singularities characterized by equation below,
\begin{equation}
\oint_ C b^\alpha_\mu dx^\mu = Z^\alpha,
\end{equation}
where $Z^\alpha$ is a non-zero integer, and the integral is evaluated around an arbitrary loop containing the defect. These singularities break local translational symmetry, invalidating the wave packet description. Our current discussion assumes explicitly that no such singularities exist.

In general, for the 1-D boson with deformation $\Omega=\Omega(x,t),\kappa=\kappa(x,t),a=a(x,t)$, the lattice vectors are $(\frac{2\pi}{\Omega},0)$ and $(\frac{\kappa\cdot a}{\Omega},a)$, and the reciprocal lattice vectors are $(\frac{\Omega}{2\pi},\frac{\kappa}{2\pi})$ and $(0,-\frac{1}{a})$. By Substituting the expression of lattice connection, we can find all the connection components are

\begin{equation}
\begin{split}
    \Gamma^{0}_{00}&=-\frac{\partial_t\Omega}{\Omega},\ \Gamma^{0}_{01}=-\frac{ \partial_x \Omega }{\Omega }, \ \Gamma^0_{10}=-\frac{a\partial_t\kappa+\kappa\partial_ta}{a\Omega}, \\
    \Gamma^0_{11}&=-\frac{a\partial_x\kappa+\kappa\partial_x a}{a\Omega},\
	\Gamma^1_{00}=0,\ \Gamma^1_{01}=0, \\
    \Gamma^1_{10}&=-\frac{\partial_t a}{a},\ \Gamma^1_{11}=-\frac{\partial_x a}{a}.
\end{split}
\end{equation}

The lattice connections above show that the connections are not torsion free because $\Gamma^\mu_{\rho \sigma}\neq\Gamma^\mu_{\sigma \rho}$. The torsion defined as the antisymmetric part of connection are $S^\mu_{\rho \sigma}=\frac{1}{2}(\Gamma^\mu_{\rho \sigma}-\Gamma^\mu_{\sigma \rho} ) $. From the definition of torsion we can obtain $S^\mu_{\rho \sigma}=-S^\mu_{\sigma \rho}$, so there are only 2 independent components of torsion, $S^0_{01}$ and $S^1_{01}$, which are
\begin{equation}
    S^0_{01}=\frac{a\partial_t\kappa+\kappa\partial_ta-a\partial_x \Omega}{2a\Omega},\quad S^1_{01}=\frac{\partial_t a}{2a}.
\end{equation}
Torsion may couple to spin and angular momentum in general relativity\cite{RevModPhys.48.393}, but since we just take one dispersion band into consideration, the coupling with spin or angular momentum may be neglected. However, we can find the torsion of connection may couple to effective gauge fields and lead to non-Abelian situations.

\section{GEODYNAMICS OF EVENT wave packet UNDER
DEFORMATION}

A free particle in curved spacetime moves along a geodesic line, following the straightest possible path determined solely by spacetime curvature \cite{carroll2019spacetime}. To indicate the motion along the path and get the pure dynamical behavior, we introduce the lattice covariant derivative, which is required to satisfy the local periodic condition for any derivative,  
\begin{equation}
    \nabla_{x^\mu}f|_{(k+2\pi b(x),x)}=\nabla_{x^\mu}f|_{(k,x)}
,\end{equation}
where $f$ is an arbitrary periodic function satisfying $f(k+2\pi b,x)= f(k,x)$, $\nabla_x$ is the covariant derivative, the expression is
\begin{equation}
    \nabla_{x^\mu}f=\partial_{x^\mu}f(k,x)-k_l \Gamma^l_{ \mu \nu}\partial_{k_\nu}f(k,x)
.\end{equation}
The expressions derived here have a similar form to those presented in Ref. \cite{PhysRevD.99.085014}. By selecting the linear and quadratic momentum functions $f=\lambda^\mu k_\mu$ and $f=\lambda^{\mu\nu} k_\mu k_\nu$, respectively, we obtain covariant derivatives for the first- and second-order expansion coefficients of the dispersion relation $\lambda$:
\begin{equation}
\begin{split}
		\nabla_{x^\sigma} \lambda^\mu = \partial_{x^\sigma} &\lambda^\mu - k_l \Gamma_{m \sigma}^l \partial_{k_m} \lambda^\mu - \Gamma^\mu_{\alpha \sigma} \lambda^\alpha \\
	    \nabla_{x^\sigma} \lambda^{\mu\nu}= \partial_{x^\sigma} \lambda^{\mu \nu}-&k_l \Gamma_{m \sigma}^l \partial_{k_m} \lambda^{\mu \nu} - \Gamma^{\mu}_{\sigma m} \lambda^{\nu m} - \Gamma^{\nu}_{\sigma m} \lambda^{\mu m}.
	\end{split}
\end{equation} 
These covariant derivatives, compared with those in general relativity, contain additional terms required to keep periodicity in phase space.

To achieve a complete phase-space representation of the equations of motion, we introduce the covariant total derivative of momentum. We require that the total derivative of a function can be expressed as:
\begin{equation}
    df(k,x) =\nabla_{x^\mu}f dx^\mu+ \partial_{k_\mu}f Dk_\mu
.\end{equation}
Therefore, the total derivative of quasi-momentum, under spacetime deformation, is defined by:
\begin{equation}
    Dk_\mu=dk_\mu+k_l \Gamma^l_{\nu \mu} dx^\nu
.\end{equation}
The total derivative thus incorporates influences from both spacetime deformation and the intrinsic periodicity in momentum space.

The Lagrangian of quasi-particle a deformed crystal can be generalized from \eqref{lag} by changing all terms to covariant forms. Following the rule above, the total derivative of $k$ is changed to $Dk$ to indicate the pure dynamic effect. As for the dispersion, besides the terms proportional to gradient  correction, there is an extra term because of the deformation. To obtain the dispersion, we turn to the dispersion operator correction first, the correction is (see supplemental material for details),
\begin{equation}
    \Delta \hat{\mathcal{L}}_{c}=\frac{1}{2}[(\hat{x}-x_c) \cdot \nabla_{x_c}\hat{\mathcal{L}}_c+ \nabla_{x_c}\hat{\mathcal{L}}_c \cdot (\hat{x}-x_c)]+\frac{1}{2}\Gamma_{\mu \sigma}^m \hat{O}_m^{\mu \sigma}
,\end{equation}
where we replace all derivatives by covariant forms, and $\hat{O}_m^{\mu \sigma}=\sum_l \frac{\partial \hat{ \mathcal{L} }\left(\left\{\tilde{R}_l-x\right\}\right)}{\partial \tilde{R}_l^m}\left(\tilde{R}_l-x\right)^\sigma\left(\tilde{R}_l-x\right)^\mu$ is the additional term from the deformation. The dispersion correction $\bra{\phi} \Delta \hat{\mathcal{L}} \ket{\phi}$ is
\begin{equation}
     \Delta \lambda=-Im[\braket{ \nabla_{x_c}u| \lambda_c-\hat{\mathcal{L}}_c| \frac{\partial u }{\partial k }}]|_{k=k_c}+
        \frac{1}{2} \Gamma_{\mu \sigma }^m \braket{\hat{O}_m^{\mu \sigma}}|_{k=k_c}
.\end{equation}
With the modified dispersion, we have the covariant Lagrangian,
\begin{equation}
	 L=-\lambda + k_{c\mu} \dot{x}_{c}^\mu + \dot{k}_{c\mu} A_{k_{c\mu}} +\dot{x}_{c}^\mu A_{{x_c}^\mu}.
\end{equation}

Using the covariant variational method $\delta f=\nabla_{x^\mu}f \delta x^\mu+ \partial_{k_\mu}f \delta_D k_\mu$, where $\delta_D k_\mu=\delta k_\mu+k_l \Gamma^l_{\nu \mu} \delta x^\nu$. The covariant equations of motion is
\begin{gather}
       \dot{ x}^\mu = \frac{\partial \lambda}{ \partial k_{ \mu} }|_{\lambda=0} - \Omega_{k_{\mu } x^\nu} \dot{x^\nu} - \Omega_{k_{\mu } k_{\nu }} \dot{k}_{\nu}, \\
        \dot{k}_{\mu} = -\nabla_{x^\mu}\lambda|_{\lambda=0} + \Omega_{x^\mu x^\nu }\dot{x^\nu}+\Omega_{x^\mu k_{\nu}} \dot{k}_{\nu}, \label{main 4}
\end{gather}
where $\dot{k}_{\mu}=\frac{Dk_{\mu}}{d\tau}$ and the equation is in the covariant form. $x$ and $k$ are the wave packet center position and momentum, we neglect lower index $c$ for simplicity.

Our framework applies generally to various quantum systems without requiring the explicit form of their action. However, for an important subset of quantum systems governed by the Schrödinger equation, the Lagrangian has a particularly simple structure. In these cases, the dispersion introduced by the operator $\hat{\mathcal{L}} = i\partial_t - \hat{H}$ always takes the form $\lambda = -\omega + \epsilon(\vec{k}, x, t)$, where $\epsilon(\vec{k}, x, t)$ is the eigenvalue of $\hat{H}$ including gradient corrections. With this dispersion relation, the equation of motion for the temporal component simplifies to:
\begin{equation}
\dot{t} = \frac{\partial \lambda}{\partial(-\omega)} = 1.
\end{equation}
The linear dependence of $\omega$ leads directly to $t = \tau$. Additionally, since the state function is independent of $\omega$, the Berry connection in $\omega$ is zero. Substituting $t = \tau$ into the Lagrangian, we obtain:
\begin{equation}
L = -\epsilon(\vec{x},\vec{k}) + \vec{k} \cdot \dot{\vec{x}} + (A\_{t} + \dot{\vec{x}} \cdot A\_{\vec{x}}) + \dot{\vec{k}} \cdot A\_{\vec{k}},
\end{equation}
which matches the known expression for wave packet dynamics in a three-dimensional deformed lattice as discussed in Refs.\cite{PhysRevB.98.115162, liangdongthesis}. Our advancement extends this scenario to explicitly include discrete time periodicity within a deformed lattice. The conventional time-independent scenario can thus be viewed as the continuum limit of this periodic structure as the time period $T \rightarrow 0$.

\section{Geodesic line and artificial gravity with deformation} 
In previous sections, we identified effective gravity induced by intrinsic dispersion of quasi-particles. However, when external deformation is introduced, emergent gravity may result from the interplay between this deformation and intrinsic dispersion of the quasi-particles. Deformation primarily determines the modification of the space-time structure, while the geometric properties of quasi-particles and dispersion relations provide higher-order corrections under the assumption of adiabatic approximation.

Following the approach established in previous sections, to account for deformation, we replace all terms with their covariant forms, analogous to the process described in equation \eqref{main 4}. Upon this substitution, we note that the commutator between the momentum derivative and the covariant derivative in coordinate space is non-zero, specifically:
\begin{equation}
\left [\nabla_{x^\nu},\partial_{k_\mu}\right ]f(k,x) = \Gamma_{\nu m}^\mu \partial_{k_m}f(k,x).
\end{equation}
Applying these covariant replacements allows us to derive the equations of motion consistent with those deformation.

First, we may find the equation of motion for a free particle. When we ignore all berry curvatures, equations \eqref{main 4} are simplified to
\begin{gather}
\begin{split}
   \dot{x}^\mu &= \frac{\partial \lambda}{ \partial k_{ \mu} } \label{main 3} \\ 
          \dot{ k}_{\mu} =& -\nabla_{x^\mu}\lambda,
\end{split}
\end{gather}
then we take an extra derivative of equation \eqref{main 3} and
\begin{equation}
    \ddot{x}^\mu-\Gamma^\mu_{\rho\sigma} \dot{x}^\rho \dot{x}^\sigma =\dot{ k}_{\nu} \partial_{k_{\mu}} \dot{x}^\nu -\dot{x}^\nu \partial_{k_{\mu}} \dot{k}_{\nu} \label{geo 1}
,\end{equation}
% change to \lambda form and give the definition of force and Lorentz force 
 The left side of equation \eqref{geo 1} is the part of covariant transport, where the lattice connection plays a similar role as the Levi-Civita symbol in general relativity, and the right side of equation \eqref{geo 1} is a rotation in phase space like we find in preview section.

The constraint defining free particles is that the covariant derivative of the dispersion should vanish,
\begin{equation}
\nabla_{x_{c}^{\mu}} \lambda(k_c,x_c)=0 \label{cons1}
.
\end{equation}
This constraint is still valid within a certain neighborhood around $x_c$ and $k_c$, leading to the additional condition $\partial_{k_{c\nu}} \nabla_{x_{c}^{\mu}} \lambda(k_c,x_c)=0$. Under this constraint, equation \eqref{geo 1} reduces to
\begin{equation}
\ddot{x}^\rho -  \Gamma^\rho_{\nu \sigma}  \dot{x}^\sigma \dot{x}^\nu = 0 \label{geo3}
,
\end{equation}
which is the standard geodesic equation. Although the entire system is embedded in a flat laboratory space-time manifold, the curved lattice structure makes quasi-particles behave as if they are moving in a curved universe. Unlike the Levi-Civita connection typically encountered in general relativity, the lattice connection described here is experimentally accessible and can be manipulated, especially in cold atom systems. The deformed lattice connection thus provides a source of first-order artificial gravity, causing observable deviations in geodesic trajectories from those predicted by the free particle geodesic equation.

When we can take the linear expansion of dispersion $\lambda=\lambda^\mu k_\mu+\lambda_0$, the geodesic equation is 
\begin{equation}
	\ddot{x}^\mu-\Gamma^\mu_{\rho\sigma} \dot{x}^\rho \dot{x}^\sigma=-\dot{x}^\nu \nabla_{x^\nu}\lambda^\mu,
\end{equation}
which tells how the acceleration depends on velocity and space. The main difference here is the covariant derivative acting on the Fermi velocity. 

Consider the dispersion up to the quadratic term, we still take the simplicity by taking a momentum shift $q_\mu=k_\mu-\boldsymbol{A}_\mu$, so that the dispersion is $\lambda=\lambda_0+\frac{1}{2} \lambda^{\mu \nu}q_\mu q_\nu$. With such form of dispersion, the geodesic line is 
\begin{equation}
\begin{aligned}
\ddot{x}^\mu-\left(\Gamma^\mu_{\rho\sigma} -\tilde {\Gamma}^\mu_{\rho\sigma} \right) \dot{x}^\rho \dot{x}^\sigma= \lambda^{\mu \nu} \partial_\nu \lambda_0-\lambda^{\mu\nu} \boldsymbol{F}_{\nu \rho} \dot{x}^\rho,\\
\tilde{ \Gamma}^\mu_{\rho\sigma}= \frac{1}{2} \lambda^{\mu \beta}(\nabla_\beta \lambda_{\rho \sigma} -\nabla_{\rho} \lambda_{\beta \sigma}- \nabla_{\sigma} \lambda_{\beta \rho}).
\end{aligned}
\end{equation}
Compared with equation \eqref{flat_geo 2}, there are two differences, the first one is we have lattice connection together with connection induced by the dispersion as an effective connection, the other is the gauge field may be non-abelian when there is torsion. The covariant gauge field is $\boldsymbol{F}_{\nu \rho}=\nabla_{\nu} \boldsymbol{A}_{\rho}-\nabla_{\rho} \boldsymbol{A}_{\nu}$, which is
\begin{equation}
    \boldsymbol{F}_{\nu \rho}=\partial_{\nu} \boldsymbol{A}_{\rho}-\partial_{\rho} \boldsymbol{A}_{\nu}+2 S^{\mu}_{\nu \rho} \boldsymbol{A}_{\mu}.
\end{equation}

The covariant equation of a free particle with berry curvatures is
\begin{equation}
\begin{aligned} 
    \ddot{x}^\rho +(\nabla_{x^\sigma} \Omega_{ k_{\rho} x^\nu } - &  \Gamma^\rho_{\nu \sigma} ) \dot{x}^\sigma \dot{x}^\nu= \\ &( \Omega_{x^l x^\alpha}\partial_{{k_{\alpha}}}\partial_{ k_{\rho}} \lambda   +  \Omega_{k_{\rho} k_{\nu}}  \nabla_{x^l}\nabla_{x^\nu} \lambda  )\dot{x}^l  \label{geo 2}
.\end{aligned}
\end{equation}
%  Generalitivity condition, the third derivative of lambda is zero. 
Equation \eqref{geo 2} is the covariant form of \eqref{flat_geo_3}. All the berry curvatures are obtained using covariant derivative to be consistent with deformation, which is also true for the new effective connection $\nabla_{x^\sigma} \Omega_{ k_{\rho} x^\nu } $ and the new metric $\nabla_{x^l}\nabla_{x^\nu} \lambda $ coupling to berry curvatures $\Omega_{k_{\rho} k_{\nu}}$ in momentum space

After obtaining the geodesic equations under various approximations and constraints, it is instructive to clarify the similarities and differences between geodesic lines in spacetime crystals and those encountered in general relativity. A key similarity is that both types of geodesic equations involve connections, highlighting the deformation of the underlying spacetime manifold. In the case of spacetime crystals, an effective connection emerges, influenced by the dispersion relations and geometric characteristics of the system, whereas in general relativity, the connection is uniquely determined by the spacetime metric itself. A crucial difference arises from the effective force term. In spacetime crystals, this effective force originates from momentum shifts and Berry curvatures, both of which are gauge fields coupled to the new dispersion metric. In summary, the effective spacetime structure experienced by quasi-particles in spacetime crystals is influenced by three factors: the deformation of the base manifold, the dispersion characteristics of quasi-particles, and the geometrical properties associated with these particles. The latter two aspects represent extensions beyond the standard formulation of general relativity.

 \section{summary and outlook} 
In this paper, we have developed a method for constructing space-time wave packets to represent events within space-time crystals and have derived the corresponding equations of motion for the particle world lines. Additionally, we have introduced a geometric framework to characterize deformation within space-time crystals, leading to a covariant formulation of the equations of motion in phase space. Detailed analysis of the geodesic equations, including higher-order corrections, revealed that the leading-order gravitational effects in space-time crystals originate from the geometry of the space-time lattice and are further modified by Berry curvature and dispersion.

A natural and intriguing extension of this work would be to generalize our method to non-Abelian cases, such as quasi-particles exhibiting spin textures or other internal degrees of freedom, which could couple explicitly to the torsion component of the lattice connection. Furthermore, addressing the Einstein field equations for the underlying space-time manifold represents a compelling direction for future research. Incorporating external dynamical elements such as domain walls and electromagnetic fields may enable explicit derivation of these equations within our framework.

\section{Appendix}
\subsection{Gradient expansion to dispersion}
Viewed in the lab frame, the total dispersion depends on space-time lattice with deformation, which is hard to deal with. Since the variation is small, we can expand the dispersion to the first order of deformation, and the zeroth order of dispersion is locally periodic in spatial and time axes. To obtain a periodic frame called a lattice frame, we introduce a smooth field in terms of lattice frame coordinates. Given the 4-positions of lattice points in lab frame $R_l$, where $l$ is a 4-component array to locate the lattice point, we define a smooth lattice field $R(x')$, which satisfies
\begin{equation}
    R_l=R(l)
,\end{equation}
and lab frame coordinates are related to lattice frame coordinates as
\begin{equation}
    x=R(x')
.\end{equation}
From the definition above, any deforming crystal is mapped to a periodic unit cell lattice in a lattice frame. The physical property of lattice vector fields comes naturally from the definition of lattice frame
\begin{equation}
    c^\mu_\alpha(\frac{l+(l+1_\alpha)}{2})=R_{l+1_{\alpha}} -R_l
.\end{equation}
Turn back to the lab frame. We can find it is consistent with the definition in the main part of the letter.

The local dispersion operator $\hat{\mathcal{L}}(k,r,x_c)$ depends on the deformation of 4-coordinates in lattice and 4-coordinates of quasi-particle. Generally, in the lab frame, we can find the spatially related part has the form like $\hat{ \mathcal{L} }(\tilde{R}_l-x)$, where $x$ is the center position of wave packet. With the lattice field $R_l=R(l)$ and $\tilde{R}_l=l^{\alpha}c_\alpha(x)$, we expand $l$ respect to the electron position in lattice frame $x'$, we have
\begin{equation}
    R_l-x=R(l)-R(x')\approx(l-x')^\alpha c_\alpha+\frac{1}{2}(l-x')^\alpha(l-x')^\beta(c_\beta \cdot \partial) c_\alpha
,\end{equation}
and the gradient correction of effective due to deformation is 
\begin{equation}
    \hat{ \mathcal{L} }(\tilde{R}_l-x)\approx\hat{ \mathcal{L} }((l^\alpha-x'^\alpha)c_\alpha(x))+\frac{1}{2}\Gamma_{\mu \sigma}^m \hat{O}_m^{\mu \sigma}
,\end{equation}
where $\hat{O}_m^{\mu \sigma}=\sum_l \frac{\partial \hat{ \mathcal{L} }\left(\left\{\tilde{R}_l-x'^\alpha c_\alpha\right\}\right)}{\partial \tilde{R}_l^m}\left(\tilde{R}_l-x'^\alpha c_\alpha\right)^\sigma\left(\tilde{R}_l-x'^\alpha c_\alpha\right)^\mu$, if we turn back to lab frame we can find the correction of deformation is $\hat{O}_m^{\mu \sigma}=\sum_l \frac{\partial \hat{ \mathcal{L} }\left(\left\{\tilde{R}_l-x\right\}\right)}{\partial \tilde{R}_l^m}\left(\tilde{R}_l-x\right)^\sigma\left(\tilde{R}_l-x\right)^\mu$, where $x$ is 4-coordinate of wave packet center in the lab frame.

Besides the deformation of lattice points, the dispersion also depends on the position of the center location of the wave packet, so use covariant derivative and equation(10) to get the gradient correction caused by the center position of the wave packet as
\begin{equation}
    \Delta \hat{\mathcal{L}}'=\frac{1}{2}[(\hat{x}-x_c) \cdot \nabla_{x_c}\hat{\mathcal{L}}_c+ \nabla_{x_c}\hat{\mathcal{L}}_c \cdot (\hat{x}-x_c)]
,\end{equation}
then the total correction is
\begin{equation}
    \Delta \hat{\mathcal{L}}_{c}=\frac{1}{2}[(\hat{x}-x_c) \cdot \nabla_{x_c}\hat{\mathcal{L}}_c+ \nabla_{x_c}\hat{\mathcal{L}}_c \cdot (\hat{x}-x_c)]+\frac{1}{2}\Gamma_{\mu \sigma}^m \hat{O}_m^{\mu \sigma}
,\end{equation}
with the total correction of effective, the approximation of eigenvalue $\lambda$ is the expectation value of correction of dispersion with a wave packet state function,
\begin{equation}
    \Delta \lambda=\braket{W|\Delta \hat{\mathcal{L}}_{c}|W}
,\end{equation}
Still according to Sundaram-Niu wave packet formalism,
\begin{equation}
     \Delta \lambda=-Im[\braket{ \nabla_{x_c}u| \lambda_c-\hat{\mathcal{L}}_c| \frac{\partial u }{\partial k }}]|_{k=k_c}+
        \frac{1}{2} \Gamma_{\mu \sigma c}^m \braket{\hat{O}_m^{\mu \sigma}}|_{k=k_c}
.\end{equation}

\color{black}
\subsection{Geodesic line equation in general}

In the main text, we obtained geodesic line equations with different corrections to find their effect, now we want a general geodesic line equation to indicate the coupling of dispersion and Berry curvatures in the most general form. By using another set of coordinates for simplicity, we rewrite the equation of motion in phase space in such a form

\begin{equation}
	M^{ij}_{\mu\nu}\dot{\xi}^{\nu}_{i}=\partial_{\xi _j^\mu}\lambda
\end{equation}
Where $M_{\mu \nu}$ is an $8\times8$ matrix 
\begin{equation}
	M^{ij}_{\mu\nu}=
	\left [ \begin{array}{cc}
		\eta_{\mu \nu}+\Omega_{k^\mu x^\nu} & \Omega_{k^\mu k^\nu} \\
		- \Omega_{x^\mu x^\nu} & \eta_{\mu \nu}-\Omega_{x^\mu k^\nu}
	\end{array} \right ],
\end{equation}
 $\xi$ is the coordinate in phase space
 \begin{equation}
 	\xi_{j}^\nu=\left [ \begin{array}{cc}
 		x^\nu \\
 		k^\nu
 	\end{array} \right ].
 \end{equation}
 The matrix and coordinate are decomposed into two parts, the top one is in real space and the bottom one is for momentum space, each of them is a 4-vector. We use the metric to make all indices of $k$ and $x$ lower indices for simplicity. The summation rule is that we sum all over the same indices in English letters $ij$ and Greek letters $\mu \nu$. The partial derivatives should follow the covariant forms
 \begin{equation}
 	\partial_{\xi _j^\mu}=\left [ \begin{array}{cc}
 	\nabla_{x^\nu} \\
 	\partial_{k^\nu}
 		
 	\end{array} \right ].
 \end{equation}
 
 Under the adiabatic approximation of deformation and perturbation, we derive the geodesic line equation in general is
 \begin{widetext}
 \begin{equation}
 	M^{ij}_{\mu\nu} \ddot{\xi}^\nu _j+\left( T^{ijk}_{\rho\mu\nu}+ \Gamma^{m;k}_{\rho\mu}A^{ij}_{m\nu}+\Gamma^{m;k}_{\rho\nu}A^{ij}_{\mu m}-\Gamma^{m;k}_{\mu\nu}M^{ij}_{\mu\rho} \right ) \dot{\xi}^{\rho}_{k}\dot{\xi}^\nu _j=\partial_{\xi^{\nu}_j}\partial_{\xi_i^\nu}\lambda \dot{\xi}_j^\nu,
 \end{equation}
 \end{widetext}
 Where $A^{ij}_{\mu\nu}$ is the Berry curvature matrix in phase space in zeroth and first order
\begin{equation}
	A^{ij}_{\mu\nu}=
	\left [ \begin{array}{cc}
		\Omega_{k^\mu x^\nu} & \Omega_{k^\mu k^\nu} \\
		0 & -\Omega_{x^\mu k^\nu}
	\end{array} \right ],
\end{equation}
and the tensor $T^{ijk}_{\rho\mu\nu}=\partial_{\xi^\rho_i}A^{jk}_{\mu\nu}$. The connection here $\Gamma^{m;k}_{\rho\sigma}$ is a vector
\begin{equation}
     \Gamma^{m;k}_{\rho\sigma} =\left [ \begin{array}{cc}
	\Gamma^{m}_{\rho\sigma} & 0
	\end{array} \right ].
\end{equation}
Since there is deformation only in real space, the component of momentum in the connection vector is zero. From the general form we can find the effect in the geometric phase is coupled to deformation in lattice and induces higher-order modification in effective connection. 
\color{black}

\subsection{An example in one-dimensional space-time crystal}
A more direct model is presented here to help illustrate our theory. Consider a 1+1D space-time crystal with arbitrary dispersion, we are going to write a geodesic line equation with parameter $t$ instead of $\tau$. The two geodesic line equations \eqref{geo3} for free particles are
\begin{align}
	\ddot{t}+\Gamma^{0}_{00}\dot{t}^2+(\Gamma^{0}_{01}+\Gamma^0_{10})\dot{x}\dot{t}+\Gamma^0_{11}\dot{x}^2=0\\
	\ddot{x}+\Gamma^{1}_{00}\dot{t}^2+(\Gamma^{1}_{01}+\Gamma^1_{10})\dot{x}\dot{t}+\Gamma^1_{11}\dot{x}^2=0.
\end{align}
\color{black}
Through the relation $\dot{x}=\dot{t}x'$ where $x'=\frac{dx}{dt}$, we have the equation of $x$
\begin{equation}
	x''+\Gamma^1_{00}+(\Gamma^{1}_{01}+\Gamma^1_{10}-\Gamma^{0}_{00})x'+(\Gamma^1_{11}-\Gamma^{0}_{01}-\Gamma^0_{10})x'^2-\Gamma^0_{11}x'^3=0,
\end{equation}
which indicates the acceleration law in the deformed space-time crystal. 

To make the simple model more detailed to indicate the true space-time structure in the geodesic line equation, we choose the one-dimensional oblique space-time crystal with Hamiltonian 
\color{black}
\begin{equation}
	\hat{H}=\sum_R [(V+A \cos ( \kappa R - \Omega t) ) \ket{R}\bra{R} + h(\ket{R+a} \bra{R}+ h.c )  ]
,\end{equation} 
where the lattice $a$, frequency $\Omega$ and $\kappa$ are all dependent in space and time. The lattice vectors are $(\frac{2\pi}{\Omega},0)$ and $(\frac{\kappa\cdot a}{\Omega},a)$. The reciprocal lattice vectors are $(\frac{\Omega}{2\pi},\frac{\kappa}{2\pi})$ and $(0,-\frac{1}{a})$. Under the geometrized unit in the main text, the lattice connection of this lattice is 
\color{black}
\begin{widetext}
\begin{align}
	&\Gamma^{0}_{00}=-\frac{\partial_t\Omega}{\Omega},\Gamma^{0}_{01}=-\frac{ \partial_x \Omega }{\Omega },\Gamma^0_{10}=-\frac{a\partial_t\kappa+\kappa\partial_ta}{a\Omega},\Gamma^0_{11}=-\frac{a\partial_x\kappa+\kappa\partial_x a}{a\Omega};\\
	&\Gamma^1_{00}=0,\Gamma^1_{01}=0,\Gamma^1_{10}=-\frac{\partial_t a}{a},\Gamma^1_{11}=-\frac{\partial_x a}{a}.
\end{align}
\end{widetext}
\color{black}
%The following equations should be met to satisfy the torsion-free condition in the main text, $\Gamma^1_{01}=\Gamma^1_{10}$ and $\Gamma^0_{01}=\Gamma^0_{10}$, which present the constraints on parameters $a,\Omega,\kappa$
%\begin{gather}
%\begin{split}
%	 \Omega\partial_x\Omega=\kappa\partial_t \Omega,\\
%	 (\partial_t-\partial_x)\ \ln(\frac{\kappa a}{\Omega})=0.
%	 \end{split}
%\end{gather}
%Notice that we are using geometrical units here, so the second equation tells us that the parameters $\frac{\kappa a}{\Omega}$ have a solution like propagating waves. The first equation is that the propagation velocity of outra-field in materials is equal to the relative changing rate of $\Omega$.

 By substituting all the connections, we can obtain the first-order geodesic line without Berry curvature. For the space part equation, it is
 \color{black}
\begin{widetext}
\begin{equation}
x''+(\frac{\partial_t\Omega}{\Omega}-\frac{\partial_t a}{a})x'+(-\frac{\partial_x a}{a}+\frac{ \partial_x \Omega }{\Omega }+\frac{a\partial_t\kappa+\kappa\partial_ta}{a\Omega})x'^2+\frac{a\partial_x\kappa+\kappa\partial_x a}{a\Omega}x'^3=0
\end{equation}
\end{widetext}
This equation is the acceleration equation in the 1+1D oblique space-time crystal. Equation (62,63) describes an oblique space-time, which has a torsion and curvature.   When $\kappa=0$, we have an orthogonal space-time, which still has a torsion.  The geodesic only depends on how the unit cell area varies in space-time.  When it only depends on time, we can solve the velocity as $v=C a / \Omega$, where $C$ is a constant.  If $a / \Omega$ is proportional to time, we obtain an expanding universe with a constant Hubble coefficient.
\color{black}
 \subsection{Gauge transformation of $\tau$}
 The parameter $\tau$ in the main text is an analogy of time, but it can be set arbitrarily since the form of $\lambda$ is not single. The requirement is that the action $S=\int d\tau \ L$ is invariant after making a transform $\tau'=f(\tau)$, which makes the action of quasi-particle changed to $S=\int d\tau' \ L'$. Considering the explicit form of the Lagrangian is $L=-\lambda + k_{c\mu} \dot{x}_{c}^\mu + \dot{k}_{c\mu} A_{k_{c\mu}} +\dot{x}_{c}^\mu A_{{x_c}^\mu}$, we find that only $\lambda$ has no first-order derivative of $\tau$, other terms are $\tau$ independent, so they are invariant under the transformation of $\tau$. To obtain the gauge transformation form of $\lambda$, we define an auxiliary field $\mu=\mu(\tau)$ and rewrite Lagrangian as
 \begin{equation}
 	L=-\mu \lambda+ k_{c\mu} \dot{x}_{c}^\mu+ \dot{k}_{c\mu} A_{k_{c\mu}} +\dot{x}_{c}^\mu A_{{x_c}^\mu}.
 \end{equation} 
  When we apply a gauge transformation $\tau'=f(\tau)$, the auxiliary field is transformed like $\mu(\tau')=\frac{d\tau}{d\tau’} \mu(\tau)$, which makes the action invariant because $\mu(\tau')d\tau=\mu(\tau)d\tau$. 
  
 The equations of motion with this auxiliary field are
\begin{gather}
 	\lambda=0, \label{app} \\
 	  \dot{ x}^\mu =\mu \frac{\partial \lambda}{ \partial k_{ \mu} }- \Omega_{k_{\mu } x^\nu} \dot{x^\nu} - \Omega_{k_{\mu } k_{\nu }} \dot{k}_{\nu}, \\
        \dot{k}_{\mu} = -\mu \nabla_{x^\mu}\lambda + \Omega_{x^\mu x^\nu }\dot{x^\nu}+\Omega_{x^\mu k_{\nu}} \dot{k}_{\nu}.
 \end{gather}
 The equation \eqref{app} is derived from the variation of `$\mu$. To simplify the set of equations we fix the gauge $\mu(\tau)=1$ in the main text and apply the constraint $\lambda=0$ in equation of motion.
\subsection{Deformation coupling to the momentum shift}
Under the expansion of dispersion near the zero of $\lambda(k)$, we may find that the dispersion is dependent on the moment after a shift of zero point $\lambda(k)=\lambda(k-A)$. The new variable $q=k-A$ is also defined in main text. If there is deformation in real space, the zero points may be space-time dependent and behave like a gauge field, which leads an additional change to the original covariant derivative $\nabla_{x^\mu}$. We may define a new covariant derivative $\tilde{\nabla}$ here,
\begin{align}
    \tilde{\nabla}_\mu&=\partial_{x^\mu}|_{q_\mu=k_\mu-A_{\mu}}-\frac{\partial A_m}{\partial x^\mu}\partial_{q_m}-k_l \Gamma_{m \sigma}^l \partial_{k_m} ,\\
    &=\partial_{x^\mu}-\left(q_l \Gamma_{m \sigma}^l \partial_{k_m}+\left(A_l \Gamma_{m \sigma}^l +\frac{\partial A_m}{\partial x^\mu} \right) \partial_{k_m} \right),\\
    &=\partial_{x^\mu}-\left(q_l \Gamma_{m \sigma}^l \partial_{k_m}+\left(A_l \Gamma_{m \sigma}^l +\frac{\partial A_m}{\partial x^\mu} \right) \partial_{q_m} \right),\\
    &=\partial_{x^\mu}-\left( q_l \Gamma^{l}_{\mu m}+\nabla_{x^\mu}A_m \right) \partial_{q_m}.
\end{align}
So we have a new covariant derivative with modification induced by gauge fields. In addition, the total derivative of momentum $q$ can be upgraded as 
\begin{align}
    \tilde{D}q_\mu=dq_\mu+\left( q_l \Gamma^{l}_{\mu m}+\nabla_{x^\mu}A_m \right) d x^m.
\end{align}

The equation of motion is still the same form, we just need to replace the variables from $(x,k)$ to $(x,p)$, the motion coupling to the effective gauge fields hides in the new covariant derivatives. By the substitution equation \eqref{main 3} is 
\begin{align}
 \dot{x}^\mu &= \frac{\partial \lambda}{ \partial k_{ \mu} } \\ 
          \dot{ q}_{\mu} =& -\tilde{\nabla}_{x^\mu}\lambda,    
\end{align}
Where $\dot{q}$ is $\frac{\tilde{D}q}{d\tau}$. When we take a step back and write it in the original covariant derivative $\nabla_{x^\mu}$ to obtain a more clear insight, we will find the equation of motion in momentum component there is a term like Lorentz force. For example, the equation \eqref{main 3} may be changed to
\begin{align}
     \dot{x}^\mu &= \frac{\partial \lambda}{ \partial q_{ \mu} } \\ 
          \dot{ q}_{\mu} =& -\nabla_{x^\mu}\lambda - \boldsymbol{F}_{\mu \rho}\dot{x}^\rho,
\end{align}
we are using the original covariant derivative here. The effective field strength $\boldsymbol{F}_{\nu \rho}$ is defined as the strength of gauge fields $\boldsymbol{F}_{\nu \rho}=\partial_{\nu} \boldsymbol{A}_{\rho}-\partial_{\rho} \boldsymbol{A}_{\nu}+2 S^{\mu}_{\nu \rho} \boldsymbol{A}_{\mu}$.
\bibliography{Main}
\end{document}